\documentclass[a4paper,fleqn]{cas-dc} 

\usepackage[numbers,square]{natbib}

\usepackage{balance}

\usepackage{xcolor}
\definecolor{revisiongreen}{RGB}{0,128,0}
\newcommand{\rev}[1]{{\color{revisiongreen}#1}}

\newtheorem{theorem}{Theorem}
\newproof{proof}{Proof}

\usepackage{tikz}

\usepackage{multirow}

\usepackage[T1]{fontenc}
\usepackage[nomath]{lmodern}
\usepackage{amsfonts}
\usepackage{amssymb}
\usepackage{booktabs}
\usepackage{array}
\usepackage{enumitem}
\usepackage{algorithm}
\usepackage[noend]{algpseudocode}
\usepackage{graphicx}
\usepackage{subcaption}
\usepackage{placeins}

\usepackage{xurl}
\hypersetup{hidelinks}
\newcommand{\Ab}{\mathbf{A}}
\newcommand{\Bb}{\mathbf{B}}
\newcommand{\Sb}{\mathbf{S}}

\newcommand{\Db}{\mathbf{D}}

\usepackage{amsopn}
\DeclareMathOperator{\diag}{diag}

\graphicspath{{./figures/}} 

\newcommand{\perm}{\operatorname{per}}
\newcommand{\E}{\mathbb{E}}
\newcommand{\R}{\mathbb{R}}

\makeatletter
\NewCommandCopy{\CASOriginalPrintEmails}{\printemails}
\RenewDocumentCommand{\printemails}{}{%
  \begingroup
  \let\@footnotetext\H@@footnotetext
  \CASOriginalPrintEmails
  \endgroup
}
\makeatother

\ExplSyntaxOn
\cs_set:Npn \__first_footerline: {}
\ExplSyntaxOff

\begin{document}

\shorttitle{The Quick and the Dead}
\shortauthors{Elbek et al.}

\title[mode=title]{The Quick and the Dead: Estimating Sparse-Matrix Permanents with Adaptive Work Filtering}

\author[1]{Deniz Elbek}
\ead{deniz.elbek@sabanciuniv.edu}
\author[1]{Yiğit Manafi}
\ead{yigit.manafi@sabanciuniv.edu}
\author[1]{Zeynep Gürdal}
\ead{zeynep.gurdal@sabanciuniv.edu}
\author[1]{Sinan Yıldırım}
\ead{sinanyildirim@sabanciuniv.edu}
\author[1]{Kamer Kaya}
\cormark[1]
\ead{kaya@sabanciuniv.edu}

\cortext[cor1]{\rev{Corresponding author.}}
\affiliation[1]{
  organization={{Sabancı University,
    Faculty of Engineering and Natural Sciences}},
  addressline={{Orta Mah., Üniversite Cad.}},
  city={{Tuzla, İstanbul}},
  postcode={{34956}},
  country={{Türkiye}}
}

\begin{abstract}
Rasmussen's permanent estimator is a simple and unbiased estimator for the permanent of a binary matrix, but its practical performance can be limited by trajectories that terminate before completing a perfect matching. These failed trajectories, together with dispersion among the surviving weights, can substantially reduce the effective sample size. Although the literature leverages techniques such as doubly-stochastic matrix scaling to improve proposal balance and support filtering to remove structurally infeasible choices, using these at every step can substantially increase the trajectory cost. Furthermore, these techniques do not directly address the choice of the next vertex whose residual neighborhood is used for the proposal. This paper uses the classical minimum-degree ordering in sparse matrix algorithms to select the next vertex with $\mathcal{O}(n + m)$ total bucket-maintenance work per trajectory, where $n$ is the number of rows/columns in the matrix and $m$ is the number of nonzeros. The proposed estimator uses adaptive schedules to invoke the more expensive scaling and filtering operations only when needed. The experiments show favorable empirical accuracy under fixed time budgets on the tested matrices.
\end{abstract}


\maketitle

\section{Introduction}

For an $n \times n$ matrix $\Ab$,
\begin{equation}
  \perm(\Ab) = \sum_{\pi \in S_n} \prod_{v=1}^n \Ab_{v,\pi(v)}\label{eq:per1}
\end{equation}
defines its permanent, where $\Ab_{v,u}$ is the $(v,u)$-th entry of $\Ab$ and $S_n$ is the set of all permutations of $\{1,\ldots,n\}$. 
For a bipartite graph, the permanent of its bipartite adjacency matrix equals the number of perfect matchings. In quantum optics, the probability amplitudes of photons are related to the permanent of a unitary matrix~\cite{aaronson11}. 
It has also been used in order statistics~\cite{balakrishnan2007} and practical applications such as DNA profiling~\cite{Narahara2013ApplicationOP}. Computing the permanent exactly is \#P-complete even for binary matrices~\cite{Valiant1979}; large dense, unstructured instances can require supercomputer-scale runs~\cite{ElbekTasyaranUcarKaya2026,LundowMarkstrom2022}. 
Hence, approximation becomes the practical alternative for larger unstructured instances.

Rasmussen proposed an unbiased, {\em sequential importance sampling} (SIS)-based
estimator for the permanent of a $(0,1)$-matrix~\cite{Rasmussen1994}, but its practical performance can be
limited by trajectories that terminate before completing a perfect matching.
Although Markov-chain approximation schemes provide polynomial-time guarantees, available implementations were reported to be impractical~\cite{Bezakova,JerrumSinclairVigoda2004,Newman}.
Many practical methods build on Rasmussen's construction, using techniques such as {\em deep rejection}~\cite{HarviainenRoyskoKoivisto2021} and {\em nested importance sampling}~\cite{HarviainenKoivisto2024}.

The literature heavily exploits two tools. First, it has been shown that {\em matrix scaling} can balance the proposal distribution and reduce estimator variance. 
Dufossé et al. use scaling to estimate the number of perfect matchings in a bipartite graph~\cite{DufosseKayaPanagiotasUcar2022}, while Harviainen et al. use it to estimate the permanent of a nonnegative matrix by nested importance sampling~\cite{HarviainenKoivisto2024}. 
In addition, to avoid dead trajectories, both studies leverage techniques to {\em filter unsupported nonzeros} whose corresponding edges belong to no perfect matching. 

Our {\bf first} contribution is to use a two-sided, dynamic, linear-time {\em{minimum-degree}}~(MD) implementation for an efficient and effective vertex selection rule inside an unbiased permanent estimator. Compared with Rasmussen’s method, this changes the order in which the residual problem is expanded to quickly reduce the risk of later dead ends. Our {\bf second} contribution is a local scaling mechanism that is faster than a full {\em{Sinkhorn--Knopp} }~(SK) iteration in practice, and our {\bf third} contribution is to schedule the more expensive scaling and filtering operations adaptively, while maintaining a high {\em perfect-matching probability}, i.e., trajectory-completion probability, and a low {\em relative spread}. Our {\bf fourth} contribution is a sharper scaling-based second-moment analysis, together with a support condition relating proposal imbalance to structural feasibility. These results motivate the use of adaptive schedules, and the experiments showed that the proposed variants better meet the empirical error targets compared to a state-of-the-art implementation DeepNIS~\cite{HarviainenKoivisto2024} and also handle larger sparse instances.

The rest of the paper is organized as follows. Section 2 introduces the background and notation. Section 3 presents the proposed estimator, including vertex selection and adaptive scaling and support-filtering policies. Section 4 reports the experimental results. Section 5 discusses related work, and Section 6 concludes the paper. Appendix~\ref{app:scheduling-bounds} provides the scaling and support bounds and their proofs.

\section{Background and Notation}

To compute the permanent of an $n \times n$ dense matrix, 
Ryser's classical formula~\cite{Ryser1963} reduces 
the naive $n!$ enumeration of~\eqref{eq:per1} to an inclusion-exclusion
sum requiring roughly $n^2 2^{n-1}$ arithmetic operations. 
Although further optimizations reduce it to ${\Theta}(n 2^{n-1})$~\cite{Wilf}, 
the complexity remains exponential. As a result, for large unstructured matrices, exact computation is infeasible, yet approximation is possible. 

For a binary matrix $\Ab\in\{0,1\}^{n\times n}$, one trajectory of Rasmussen's estimator~\cite{Rasmussen1994} 
processes the rows in a fixed order and matches row $i$ with a column 
$j \in \cal{C}$, the set of unmatched columns.
From a single trajectory, a sample $X$ is drawn
with the  procedure {given in Algorithm\ \ref{alg:rasmussen}}.

\begin{algorithm}[H]
\footnotesize
\caption{Rasmussen's permanent estimator}
\label{alg:rasmussen}
\begin{algorithmic}[1]
\Require A binary matrix $\Ab\in \{0,1\}^{n\times n}$
\Ensure One trajectory contribution $X$
\State ${\cal C}\leftarrow\{1,\ldots,n\}$ and $X\leftarrow 1$
\For{$v=1,\ldots,n$}
  \State $\mathcal{N}_v\leftarrow\{u\in {\cal C}:\Ab_{v,u}=1\}$
  \If{$\mathcal{N}_v=\emptyset$}
    \Return $0$
  \EndIf
  \State $X\leftarrow X \times |\mathcal{N}_v|$
  \State Choose $u$ uniformly at random from $\mathcal{N}_v$
  \State ${\cal C}\leftarrow {\cal C}\setminus\{u\}$
\EndFor
\Return $X$
\end{algorithmic}
\end{algorithm}

Repeating this procedure independently gives samples $X_1,\ldots,X_M$; their
mean $\overline X_M = M^{-1}\sum_{k=1}^M X_k$ estimates $\mu=\perm(\Ab)$. A completed perfect matching $\pi$ is sampled with probability $p(\pi)$ and returns $X=1/p(\pi)$, so its contribution to $\E[X]$ is one. A failed trajectory terminates without a perfect matching and returns zero. Thus $\E[X]=\perm(\Ab)$.

The concentration of the sample mean is important for quickly reaching the desired 
error threshold. Assume $\mu > 0$ and let 
$\sigma^2=\operatorname{Var}(X)$, and
$\operatorname{CV}^2=\sigma^2/\mu^2$, where CV denotes the coefficient of variation. Since
$0\leq X\leq n!$, a variance-sensitive Bernstein bound gives
\begin{equation*}
  \Pr\!\left[
    \left|\frac{\overline X_M}{\mu}-1\right|\geq\epsilon
  \right]
  \leq
  2\exp\!\left(
    -\frac{M\epsilon^2}
    {2\operatorname{CV}^2+2n!\epsilon/(3\mu)}
  \right).
\end{equation*}
Thus reducing the relative variance directly improves concentration, especially when
the boundedness term does not dominate. We define the effective sample size (ESS) and an estimator of it as
\begin{equation*}
  M_{\mathrm{eff}}
  =M\frac{\mu^2}{\E[X^2]}
  =\frac{M}{1+\operatorname{CV}^2},
  \qquad
  \widehat M_{\mathrm{eff}}
  =\frac{(\sum_{k=1}^M X_k)^2}{\sum_{k=1}^M X_k^2}.
\end{equation*}
In particular,
$\operatorname{Var}(\overline X_M)/\mu^2
=1/M_{\mathrm{eff}}-1/M$.  The perfect-matching probability
$p_{\mathrm{match}}=\Pr[X>0]$ imposes an additional limit. By
Cauchy--Schwarz, $\frac{M_{\mathrm{eff}}}{M}\leq p_{\mathrm{match}}$ and
$\operatorname{CV}^2\geq p_{\mathrm{match}}^{-1}-1$.  Hence, failed
trajectories alone can collapse the effective sample size, while dispersion
among the positive trajectories can reduce it further. For this reason, we focus on both 
the perfect-matching probability and the relative spread.

\paragraph{Support and total support.}
A nonnegative square matrix $\Ab\in\R^{n\times n}$ \emph{has support} if
there exists $\pi\in S_n$
such that $\Ab_{v,\pi(v)}>0$ for every $v$. For a nonnegative matrix, this is
also equivalent to $\perm(\Ab)>0$. The stronger property of \emph{total support} 
requires that for every positive entry $\Ab_{v,u}$ there exists a 
permutation $\pi\in S_n$ such that $\pi(v)=u$ and $\Ab_{i,\pi(i)}>0$ for every $i=1,\ldots,n$. Thus, a matrix can have support but not total support.

Let $G_{\Ab}=(V_R\cup V_C,E)$ be the corresponding, undirected bipartite graph of
$\Ab$: row vertex $r_i$ is adjacent to column vertex $c_j$
if and only if $\Ab_{ij}>0$. In $G_{\Ab}$, support means that at least one
perfect matching exists, whereas total support means that every edge belongs
to at least one perfect matching. 

\paragraph{Matrix scaling.}
The Sinkhorn--Knopp~(SK) iteration alternately normalizes the rows and columns of a
nonnegative matrix; when $\Ab$, with $m$ nonzeros has {\em total support}, the iterations converge to a
doubly stochastic matrix~\cite{SinkhornKnopp1967}. 
The complexity of a single SK iteration is $\Theta(m)$ 
and the number of iterations required for a small error in row/column sums is usually small in practice.

Matrix scaling has a strong connection to the permanent.
{Let $\Sb = \Db_R\Ab\Db_C$ be a doubly stochastic scaling of $\Ab \in \{0,1\}^{n \times n}$, where $\Db_R = \operatorname{diag}(\mathbf{r})$ and $\Db_C = \operatorname{diag}(\mathbf{c})$ are diagonal row- and column-scaling matrices. Since $\perm(\Sb)=\perm(\Ab)\prod_v {\mathbf r}_v \prod_u{\mathbf c}_u$,} for a supported $(v,u)$,
\begin{equation}
  {\mathbf r}_v {\mathbf c}_u =
  \frac{\perm(\Ab \setminus \{v,u\})}{\perm(\Ab)}\,
  \frac{\perm(\Sb)}{\perm(\Sb \setminus \{v, u\})}.\label{eq:scaled-perm-ratio}
\end{equation}
Thus the value ${\mathbf r}_v {\mathbf c}_u$  is not only useful for balancing; it is connected to the permanent of $\Ab \setminus \{v,u\}$, the {\em child} matrix obtained by deleting row $v$ and column $u$.

\paragraph{Support filtering with Dulmage--Mendelsohn decomposition.}
An edge~(and the associated nonzero) is {\em{supported}} if it belongs to at least one perfect
matching of $G_{\Ab}$.  Since every nonzero product in~\eqref{eq:per1} implies a perfect matching of
$G_{\Ab}$, deleting nonzeros associated with unsupported edges leaves
$\perm(\Ab)$ unchanged.

The fine-grained Dulmage--Mendelsohn~(DM) decomposition identifies supported nonzeros~\cite{DulmageMendelsohn1958}. It first computes a
perfect matching $\mathcal M$; if none exists, then $\perm(\Ab) = 0$.  Otherwise,
orient every edge in $\mathcal M$ from $V_R$ to $V_C$ and every edge in
$E\setminus\mathcal M$ from $V_C$ to $V_R$.  Every edge of $\mathcal M$ is
supported. For an edge outside $\mathcal M$, the alternating-cycle
characterization implies that it is supported if and only if its endpoints lie
in the same strongly connected component~(SCC) of the directed graph. All
other nonzero entries may be removed. A perfect matching can be found in
$\mathcal{O}(m\sqrt n)$ time~\cite{HopcroftKarp1973}, and the SCCs require $\mathcal{O}(m+n)$ time~\cite{Tarjan1972}, where $m =|E|$. 

\section{An Adaptive Estimator for Nonnegative Matrices}

The original Rasmussen estimator fixes the row order and uses a uniform neighbor proposal. The literature primarily focuses on the proposal while retaining a prescribed row order, or orders the remaining vertices from scratch without maintaining it incrementally. We instead treat vertex selection and neighbor sampling as
equally important. Since the proposal quality depends on scaling and nonzero filtering, we also schedule these operations within the current trajectory.  
Algorithm~\ref{alg:adaptive-sis} combines the three pieces: ({\bf{1}}) a vertex selection policy $\mathcal{R}_V$, ({\bf{2}}) a scaling policy $\cal{P}_{SK}$, and ({\bf{3}}) a support-filtering policy $\cal{P}_{DM}$, while adapting the total work to the earlier decisions in the trajectory.

\begin{algorithm}[htbp]
\renewcommand{\arraystretch}{0.94}
\footnotesize
\caption{Generic permanent estimator}
\label{alg:adaptive-sis}
\begin{algorithmic}[1]
\Require{\ \newline
$\Ab\in\R^{n\times n}$, a nonnegative sparse matrix.\newline 
Vertex selection rule $\mathcal{R}_V$.\newline
Outer SK iterations $K_0$; inner SK iteration cap $K$.\newline
SK-policy $\mathcal{P}_{SK} \in \{\textsc{Fixed},\textsc{Adaptive},\textsc{Local}\}$.\newline
DM-policy $\mathcal{P}_{DM}\in \{\textsc{All},\textsc{Thrs}(\theta_{\rm th}),\textsc{Trck}(\theta_{\rm tr})\}$.
}
\Ensure{One trajectory contribution $X$}\vspace*{1ex}
\Statex{\bf{\hspace*{-3ex}Preprocessing: \textnormal{(reused by all trajectories)}}}
\State $(\Bb^{(0)},ok)\leftarrow\operatorname{InitialDM}(\Ab)$
\label{line:dm0}
\If{$\neg ok$}
  \Return $0$
\EndIf
\State Initialize scaling vectors $\mathbf{r}^{(0)},\mathbf{c}^{(0)}\leftarrow\mathbf{1}$ and MD buckets $\mathcal Q_0,\ldots,\mathcal Q_n$ (Algorithm~\ref{alg:mindeg})
\State Apply $K_0$ weighted SK iterations to $\Bb^{(0)}$ \label{line:sk0}
\vspace*{1ex}\Statex{\hspace*{-3ex}{\bf{One trajectory:}} \textnormal{(independent across runs)}}
\State {Restore a fresh copy of the initial residual, initial scaling vectors and minimum-degree buckets}
\State $X\leftarrow 1$
\State $track_{\rm dm}\leftarrow 1$
\For{$t=0,\ldots,n-1$}
  \State Choose an active row $v=\mathcal{R}_V(\Bb^{(t)})$\label{line:vertex}
  \Statex \ \ \ \ \ $\blacktriangleright$ {\tt column case is symmetric}
  \If{$\mathcal N_t(v)=\emptyset$}\label{line:deadend}
    \Return $0$
  \EndIf
  \State Construct $q_t(\,\cdot\mid v)$ from the scores in
  Eq.~\eqref{eq:scaled-base-proposal}
  \State Draw $u\sim q_t(\,\cdot\mid v)$\label{line:proposal}
  \State $z^{(t)}\leftarrow \mathbf{r}^{(t)}_{v} \mathbf{c}^{(t)}_{u}$
  \State $X\leftarrow X \times \Bb^{(t)}_{v,u}/q_t(u\mid v)$\label{line:correction}
  \State $\Bb^{(t+1)}\leftarrow\Bb^{(t)}\setminus\{v,u\}$
  \State {Update the MD buckets for the edges deactivated}
  \State {$track_{\rm dm} \leftarrow track_{\rm dm} \times z^{(t)}$}
  \If{$t+1<n$}
    \State $run_{\rm dm}\leftarrow\textsc{False}$
    \If{$\mathcal{P}_{DM}=\textsc{All}$ {\bf{ or}} \\
    \ \ \ \ \ \ \ \ \ \ \ \ \ $\mathcal{P}_{DM}=\textsc{Thrs}\land z^{(t)} <\theta_{\rm th}$  {\bf{ or}}\\
    \ \ \ \ \ \ \ \ \ \ \ \  \ $\mathcal{P}_{DM}=\textsc{Trck}\land track_{\rm dm}<\theta_{\rm tr}$}
      \State $run_{\rm dm}\leftarrow\textsc{True}$
    \EndIf
    \If{$run_{\rm dm}$}
      \State $(\Bb^{(t+1)},ok)\leftarrow
      \operatorname{InnerDM}(\Bb^{(t+1)})$
      \If{$\neg ok$}
        \Return $0$
      \EndIf
      \State {Update the MD buckets for edges deleted by DM}
      \State $track_{\rm dm}\leftarrow 1$
      \label{line:post-dm-sk}
    \EndIf
    \If{$\mathcal Q_0\ne\emptyset$}
      \State \Return $0$
    \EndIf
    \State Scale to $\Bb^{(t+1)}$ w.r.t. $\mathcal{P}_{SK}$  \label{line:child-sk}
  \EndIf
\EndFor
\State{\bf{return}} $X$
\end{algorithmic}
\end{algorithm}

During a trajectory in Algorithm~\ref{alg:adaptive-sis}, let $\Bb^{(t)}$ be the
{\em residual} matrix after $t$ matched pairs have been removed. Similarly, let $\mathcal N_t(v)$ be the active neighbors of a vertex $v$~(row or column). Assume that $v$ corresponds to a row; the column case is symmetric.
Let $\Bb^{(t)}_{v,u}$ be the matrix entry associated with edge $(v,u)$. The one-step correction is $C_t(v,u)=\Bb^{(t)}_{v,u}/q_t(u\mid v)$ where the  scores used by the proposal are computed as
\begin{equation}
  q_t(u\mid v)=
  \frac{s_t(v,u)}{\sum_{\ell\in\mathcal N_t(v)}s_t(v,\ell)}
  \label{eq:scaled-base-proposal}
\end{equation}
where $s_t(v,u)=\Bb^{(t)}_{v,u}{\mathbf c}^{(t)}_u$ 
for a selected row $v$ and a candidate column $u$. Setting all scaling vectors to one gives the unscaled weighted proposal, equivalent to Rasmussen's proposal for binary matrices with $\Bb^{(t)}_{v,u} = 1$. Note that the residual weights $\Bb^{(t)}$ are never overwritten by SK. For efficiency, an SK iteration updates only the scaling vectors $\mathbf{r}^{(t)}$ and $\mathbf{c}^{(t)}$. Hence, line~\ref{line:correction} uses the actual entry. For a column $v$, the definitions are similar.

The ideal sampler would sample each perfect matching in proportion to 
its permanent contribution. For a binary matrix, each perfect matching should therefore be sampled uniformly. Hence, the ideal proposal for both Algs.~\ref{alg:rasmussen} and~\ref{alg:adaptive-sis} is \begin{equation}
\label{eq:idealpropoisal}
q_t(u | v) = \frac{\Bb^{(t)}_{v,u}\times\perm(\Bb^{(t)} \setminus \{v,u\})}{\perm(\Bb^{(t)})}.
\end{equation}
With this proposal, each trajectory contribution is the permanent, and the variance is zero. Nevertheless, this equation cannot be evaluated efficiently: its
numerator contains exactly the permanents of the candidate residual matrices
that we are trying to estimate. This is the central circularity:
the ideal proposal solves the estimation problem.
Fortunately, as discussed below, scaling supplies a computable proxy. 

\paragraph{Unbiasedness under adaptive policies.}
Suppose that each proposal is positive on every retained nonzero and that vertex selection uses only the trajectory history. Let $W_t$ be the accumulated importance correction before step $t$, and let $\mathcal H_t$ denote that history. Permanent expansion along the selected row gives
\begin{equation*}
\begin{aligned}
 &\E[W_{t+1}\perm(\Bb^{(t+1)})\mid\mathcal H_t]\\
 &\quad=W_t\sum_{u\in\mathcal N_t(v)}
 \Bb^{(t)}_{v,u}\perm(\Bb^{(t)}\setminus\{v,u\})
 =W_t\perm(\Bb^{(t)}).
\end{aligned}
\end{equation*}
The column case is identical. DM preserves child permanents, and scaling changes only the proposal. With $\perm(\emptyset)=1$ and zero contributions from failed trajectories, induction over at most $n$ steps proves $\E[X]=\perm(\Ab)$. Thus an independent, fixed-size sample mean is unbiased.

\subsection{Outer scaling}
\begin{figure}
\centering
    \begin{subfigure}[b]{0.49\linewidth}            
            \includegraphics[width=\textwidth]{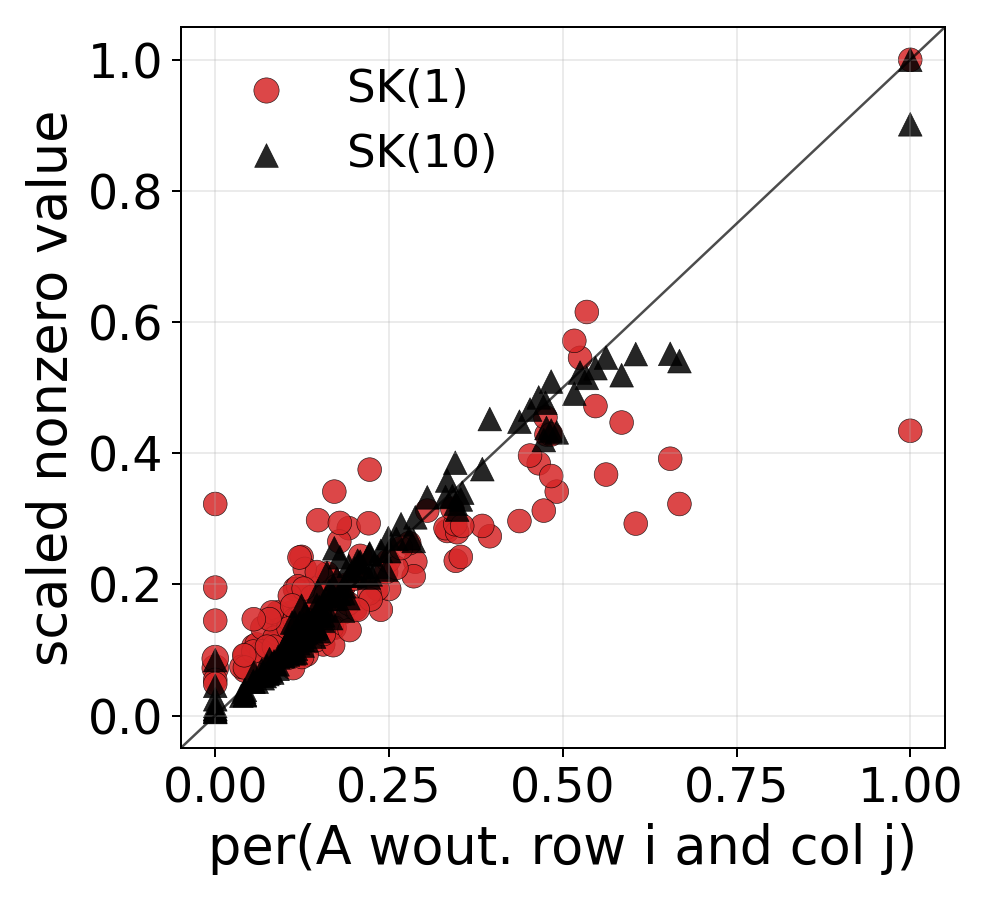}
            \caption{\small{Uniform, fixed set size\\
            ($n=30$, $m=145$).}}
            \label{fig:scaling-ex5}
    \end{subfigure}
    \begin{subfigure}[b]{0.49\linewidth}
            \centering
            \includegraphics[width=\textwidth]{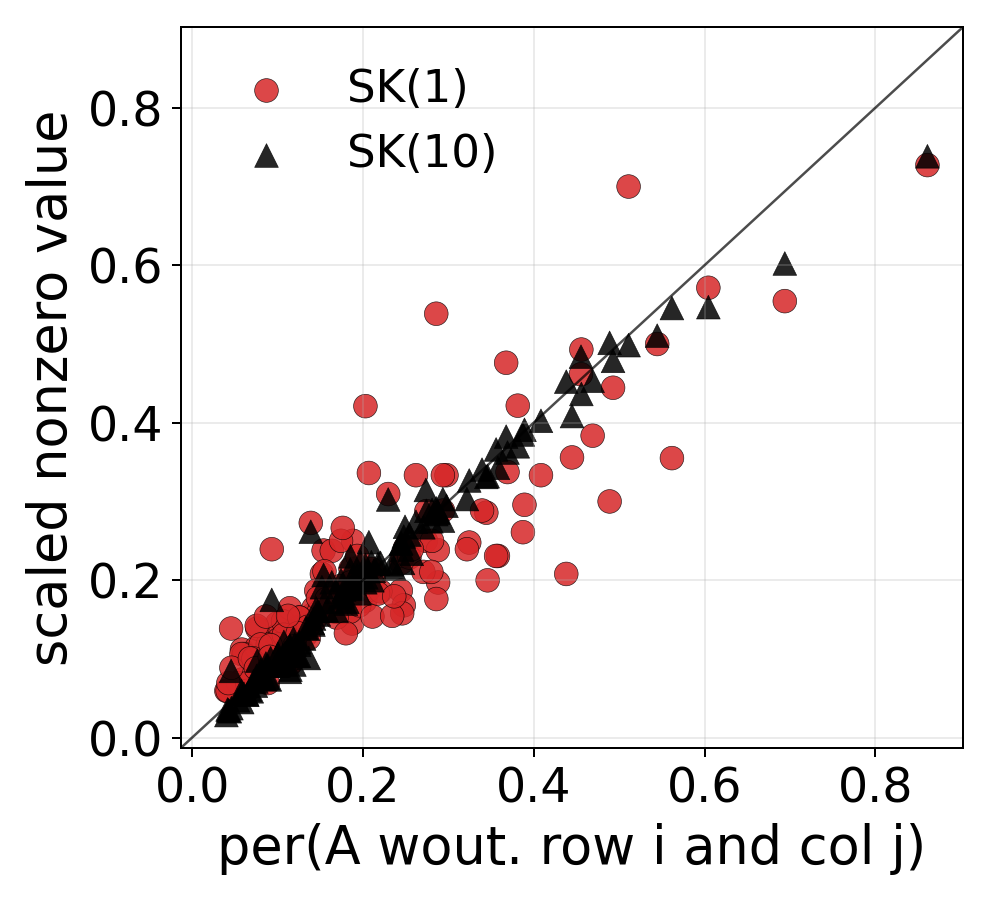}
            \caption{\small{Bernoulli, independent
            ($n=30$, $m=143$).}}
            \label{fig:scaling-bernoulli}
    \end{subfigure}
    \caption{\small{Numerically scaled nonzero values after one and ten SK
    iterations versus exact child-permanent ratios. The diagonal represents equality of the exact child-permanent ratio and the scaled value.}}\label{fig:scaling-minor-proxy}
\end{figure}

Although scaling is not guaranteed to be ideal~(Eq.~\eqref{eq:scaled-perm-ratio}),
Figure~\ref{fig:scaling-minor-proxy} tests directly whether it recovers its
relative masses.  For every supported nonzero $(v,u)$, the horizontal coordinate is
$\frac{\perm(\Ab \setminus \{v,u\})}{\perm(\Ab)}$, the exact probability that $(v,u)$ occurs in a
uniformly drawn perfect matching, and the vertical
coordinate is its scaled value $r_v c_u$.  The two test matrices are $30\times30$ binary 
matrices. For the \texttt{Uniform}($30,4$) instance, we sample exactly $4\times30$ distinct positions uniformly without replacement from the full matrix, add the full diagonal, and apply DM to obtain total support. In total, there are $m = 145$ nonzeros.
The {\tt{Bernoulli}}($30$, $4$) instance samples every position independently with
a probability of $4/30$ before adding the main diagonal, resulting in 143 nonzeros.

With SK(0), the scaling values equal one.  One  SK iteration already
moves the values toward the ideal, as suggested by
Eq.~\eqref{eq:scaled-perm-ratio}; 10 iterations make the relation substantially
tighter. Over all nonzeros, the Pearson correlation increases from
$0.85$ to $0.99$ on {\tt{Uniform}}, and from $0.87$ to $0.99$ on {\tt{Bernoulli}}. The results indicate that SK supplies a useful proposal proxy.

The generic algorithm performs an unconditional initial DM once, and 
all trajectories use the resulting matrix with total support; if no perfect matching is found at this stage, the permanent is zero. The outer SK phase~(line \ref{line:sk0}) initializes the  scaling vectors $\mathbf{r}^{(0)}$ and $\mathbf{c}^{(0)}$ corresponding to the diagonal scaling matrices
$\Db_R^{(0)}=\diag(\mathbf{r}^{(0)})$ and $\Db_C^{(0)}=\diag(\mathbf{c}^{(0)})$. 
Within a trajectory, every full or local SK operation updates these vectors, and the residual matrix weights are used without being overwritten. Although scalings align better for a binary matrix in theory, practical experiments show that it is also effective for nonnegative matrices.

\begin{figure*}[pos=t]
\color{revisiongreen}
\centering
    \begin{subfigure}[t]{0.95\textwidth}
        \centering
        \includegraphics[width=\linewidth]{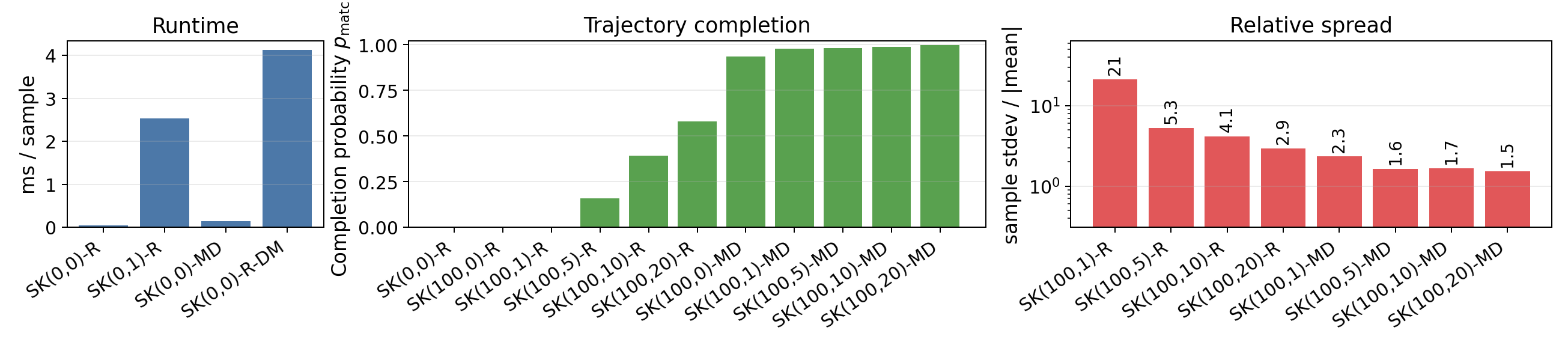}
        \caption{\texttt{sphere3}: $n=258$ and $m=1{,}026$.}
        \label{fig:md-toolstudy-sphere3}
    \end{subfigure}

    \vspace{0.4em}
    \begin{subfigure}[t]{0.92\textwidth}
        \centering
        \includegraphics[width=\linewidth]{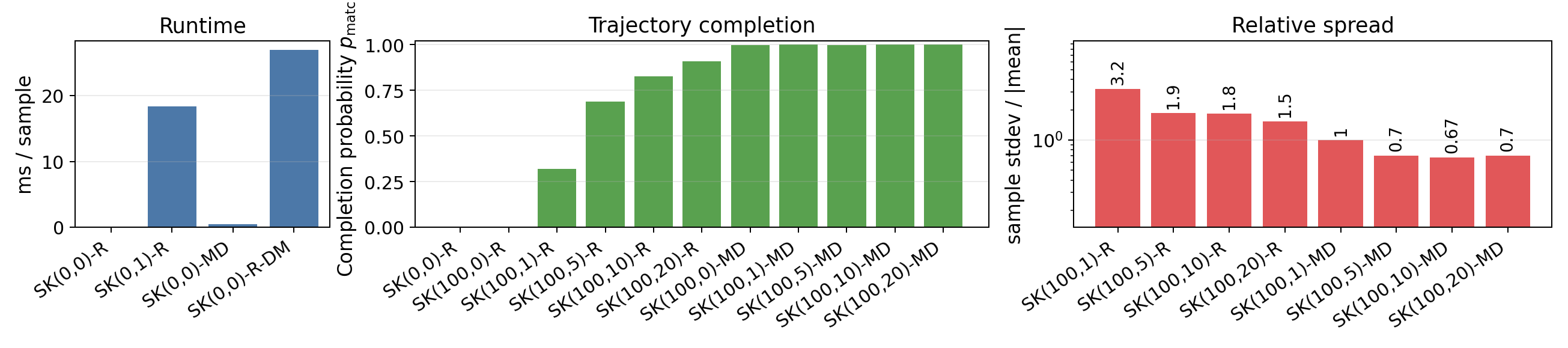}
        \caption{\texttt{Trefethen\_500}: $n=500$ and $m=8{,}478$.}
        \label{fig:md-toolstudy-trefethen}
    \end{subfigure}
    \caption{\small The time per sample~(left),  $p_{match}$~(middle), and relative spread~(right) with and without the minimum-degree policy and different numbers of outer and inner SK iterations, and no inner DM, using 1,000 independent trajectories per variant.
    SK($K_0$,$K$)-R uses fixed row order, and SK($K_0$,$K$)-MD uses incremental two-sided minimum degree; $K_0$ is the one-time outer iteration count, and $K$ is the number of full inner iterations after each choice.  The runtime figure~(left) isolates unscaled sampling, one inner SK iteration, MD bucket maintenance, and all-step DM.}
    \label{fig:md-toolstudy}
\end{figure*}

\subsection{Choosing the next vertex}

At every trajectory step, an active vertex is selected to be matched. A deterministic active vertex selection rule $\mathcal{R}_V$ introduces no
additional random factor and therefore requires no correction, even when its
choice depends on the trajectory history.  For the drawn neighbor $u$,
multiplication of the contribution by
$\Bb^{(t)}(v,u)/q_t(u\mid v)$ cancels its proposal probability in
conditional expectation.  Unbiasedness only requires $q_t$ to be positive for
every nonzero whose child residual has a positive permanent. In their experiments,
Dufoss\'e~et~al. exploit this freedom using a row-side minimum-degree rule,
but do not consider column-side candidates
\cite{DufosseKayaPanagiotasUcar2022}.  Their implementation recomputes the
candidate row degrees from the residual matrix at each of the $n$ sampling
depths.  Hence, with $m=\operatorname{nnz}(\Ab)$, number of nonzero elements in $\Ab$, vertex selection alone has
worst-case $\mathcal{O}(nm)$ overhead per trajectory. This overhead is consequential
because it limits the number of trajectories sampled per second.

\paragraph{{Two-sided, linear-time minimum degree.}}
To find a greedy matching with high cardinality, processing a low-degree vertex early is a proven engineering decision to reduce the chance that later choices consume all of its options~\cite{duku:12m, klmu:13m}.
To avoid repeated scans of the maintained degrees, which yield $\mathcal{O}({n^2 + m})$ time complexity, we maintain both bipartite sides in buckets $\mathcal Q_0,\ldots,\mathcal Q_n$. Each active vertex $x$ stores its degree $d(x)=|\mathcal N_t(x)|$ and its position in the bucket array, allowing constant-time deletion and adjacent-bucket migration. Algorithm~\ref{alg:mindeg} gives the maintenance operations; bucket-head selection resolves ties deterministically. Both adjacency directions are stored, and each active edge is deactivated only once. Selection is called only while active vertices remain; a degree-zero vertex returns a zero trajectory contribution.

\begin{algorithm}[htbp]
\footnotesize
\caption{Incremental 2-sided min-degree rule}
\label{alg:mindeg}
\begin{algorithmic}[1]
\State Initialize $d(x)=|\mathcal N_0(x)|$, buckets $\mathcal Q_{d(x)}$, and $d_{\min}\leftarrow0$
\Procedure{SelectVertex}{}
  \While{$\mathcal Q_{d_{\min}}=\emptyset$}
    \State $d_{\min}\leftarrow d_{\min}+1$
  \EndWhile
  \State \Return the head of $\mathcal Q_{d_{\min}}$
\EndProcedure
\Procedure{DeleteEdge}{$e=(x,y)$}
  \State Deactivate $e$
  \For{each still-active endpoint $w\in\{x,y\}$}
    \State Remove $w$ from $\mathcal Q_{d(w)}$; set $d(w)\leftarrow d(w)-1$
    \State Insert $w$ into $\mathcal Q_{d(w)}$
    \State $d_{\min}\leftarrow\min\{d_{\min},d(w)\}$
  \EndFor
\EndProcedure
\State On matching $(v,u)$: remove both endpoints from their buckets, mark both inactive, and call \textsc{DeleteEdge} for each active incident edge
\State After DM: call \textsc{DeleteEdge} for each deleted active edge
\end{algorithmic}
\end{algorithm}

\emph{Complexity.} Initialization, including restoration of the initial state, costs $\mathcal O(n+m)$. Each edge is deactivated at most once and causes at most two constant-time degree decrements. Each decrement lowers $d_{\min}$ by at most one; its total downward movement is therefore at most $2m$, and its total upward movement is at most $n+2m$. Thus all vertex selections and bucket updates require $\mathcal O(n+m)$ total time per trajectory and $\mathcal O(n+m)$ space. Note that we never rebuild the degrees from scratch after an inner DM call.

Figure~\ref{fig:md-toolstudy} shows how the vertex selection rule and outer/inner SK-iteration counts affect runtime, the perfect-matching probability, and relative spread for two matrices: {\tt sphere3} and {\tt Trefethen\_500}. Formally, for $M$ trajectories, we report the empirical 
$p_{\rm match}=M^{-1}\sum_{k=1}^M\mathbf{1}_{\{X_k>0\}}$ and the
\emph{relative spread}
\begin{equation*}
  {\operatorname{\widehat{CV}}}=
  \frac{{\operatorname{sd}}(X_1,\ldots,X_M)}
       {|\overline X_M|}.
\end{equation*}
By the Cauchy--Schwarz relation above,
$M_{\rm eff}/M\leq p_{\rm match}$, so ordering and support filtering can increase $p_{\rm match}$ and raise this ceiling on effective sample size. 

With a fixed row
order, neither unscaled sampling nor 100 outer SK iterations alone produces a
single complete trajectory on either matrix.  Replacing only the vertex rule
by MD~(minimum degree, distinct from DM support filtering) after the same outer initialization raises the perfect-matching probability
to $0.937$ on \texttt{sphere3} and $0.998$ on
\texttt{Trefethen\_500}. At ten inner iterations, the fixed-order and MD
probabilities are, respectively, $0.39$ and $0.99$ on \texttt{sphere3}, and
$0.83$ and $1.00$ on \texttt{Trefethen\_500}. Accordingly, the corresponding relative spreads decrease from $4.14$ to $1.68$ and from $1.84$ to $0.67$.
These substantial gains come at very low overhead thanks to the incrementally maintained minimum-degree state~(Figure~\ref{fig:md-toolstudy}-left). MD bucket maintenance costs only $0.151$ and $0.471$ ms per trajectory on the two matrices, substantially less than the reported SK or DM work. Recomputing all degrees at every depth would forfeit this advantage under a fixed-time budget.

\subsection{Scheduling Scaling and Filtering}

Scaling and DM target different terms in the ideal proposal.  SK changes the
relative proposal masses and can reduce dispersion among positive trajectory
weights.  DM removes exactly those nonzeros whose child permanent is zero and can
therefore increase the perfect-matching probability.  Neither operation changes
the permanent: SK is used only to construct $q_t$, with the
importance correction retained~(line~\ref{line:correction}), and DM removes
only unsupported nonzeros.  Their costs, however, motivate
the following independent schedules.

\paragraph{Why scale products guide the schedules.}
For an exactly doubly stochastic scaling of a binary residual of order $k$, deleting a selected pair with $z=\mathbf r_v\mathbf c_u$ leaves total scaled mass $k-2+z$. The total row deficit, and likewise the column deficit, is therefore $1-z$. Smaller $z$ means a larger loss of balance and motivates more SK work. Appendix~\ref{app:scheduling-bounds} sharpens the scaling bound of~\cite{DufosseKayaPanagiotasUcar2022} by retaining the remaining row and column masses.

Let
$\mathbf Q_{i,j}=q_t(j\mid i)$ be the row-normalized proposal
matrix from~\eqref{eq:scaled-base-proposal}, and define $
 \Delta=\tfrac12\|\mathbf Q^T\mathbf1-\mathbf1\|_1.$
The following theorem proves that if $\mathbf Q_{v,u}>\Delta$, the child has a positive permanent and helps to explain the reasoning behind our DM trigger: 

\begin{theorem}[Support criterion]
\label{thm:proposal-support}
Let $\mathbf Q\in\mathbb R_{\geq0}^{n\times n}$, $n\geq2$,
be row stochastic, so that
$\sum_{j=1}^n\mathbf Q_{ij}=1$ for every row $i$.
Define
\[
 \lambda_j=\sum_{i=1}^n\mathbf Q_{ij},
 \qquad
 \Delta=\sum_{j=1}^n(\lambda_j-1)_+
       =\frac12\sum_{j=1}^n|\lambda_j-1|,
\]
where $(x)_+=\max\{x,0\}$.

If $\mathbf Q_{vu}>\Delta$, then the residual matrix obtained by
deleting row $v$ and column $u$ has a positive permanent.
Equivalently, $(v,u)$ belongs to a perfect matching of
the positive entries of $\mathbf Q$.
\end{theorem}

\begin{proof}
Suppose that the residual has
no perfect matching. By Hall's theorem, some set of
remaining rows $R$ has a set of neighboring columns $C$
satisfying $|C|<|R|$. All indices refer to the original
matrix.

Set $W=C\cup\{u\}$. Since $u\notin C$,
\[
 |W|=|C|+1\leq |R|.
\]
The only deleted column is $u$, so every positive entry
of the original rows $R$ lies in columns $W$.
Row normalization therefore gives
\[
 \sum_{i\in R}\sum_{j\in W}\mathbf Q_{ij}=|R|.
\]
Moreover, $v\notin R$, and $\mathbf Q_{vu}$ contributes
an additional amount to these columns. Thus
\[
 \sum_{j\in W}\lambda_j
 =\sum_i\sum_{j\in W}\mathbf Q_{ij}
 \geq |R|+\mathbf Q_{vu}.
\]
Consequently,
\[
 \begin{aligned}
 \Delta
 &\geq \sum_{j\in W}(\lambda_j-1)
 &\geq |R|+\mathbf Q_{vu}-|W|
 &\geq \mathbf Q_{vu}.
 \end{aligned}
\]
This contradicts $\mathbf Q_{vu}>\Delta$.
Hence, the child has a perfect matching and a positive
permanent.
\hfill$\square$
\end{proof}

As an interesting consequence, the theorem can be used as a stopping test for support certification during SK scaling. After row normalization at iteration $i$, let
$\mathbf Q^{(i)}$ be the current matrix and $\Delta_i$ its column
imbalance. If
$\min_{v,u:\,\mathbf Q^{(i)}_{vu}>0}\mathbf Q^{(i)}_{vu}>\Delta_i$,
then every positive entry belongs to a perfect matching, so the
current residual has total support and requires no DM filtering.
For an input with total support, $\Delta_i$ tends to zero while the positive entries
approach positive limits, so the test eventually succeeds.
At the doubly stochastic limit, the same support property follows
from the Birkhoff--von Neumann representation
$\mathbf S=\sum_\ell\alpha_\ell\mathbf P_\ell$: each entry is the
total coefficient weight of the permutation matrices containing
it~\cite[Eq.~(1)]{DUFOSSE201868}.

On the contrary, if the input has a positive permanent but contains unsupported
entries, these remain positive at every finite exact iteration
and vanish only in the limit~\cite{SinkhornKnopp1967}.
The test above therefore cannot succeed until
such entries are removed. 

For efficiency, our tested policies below use raw $z$ without computing $\Delta$.
Therefore, the fixed thresholds used in our implementation remain as heuristic decisions. However, following the theorem and consequent observations, we are planning to work on more advanced scheduling approaches that maintain or approximate $\Delta$.

\paragraph{SK scheduling policies (${\cal P}_{SK}$):}
An outer phase of $K_0$ alternating row/column normalizations initializes the scaling vectors
$\mathbf{r}^{(0)}$ and $\mathbf{c}^{(0)}$. When applied while following a random trajectory, an SK iteration first updates
every active column and then every active row:
\begin{equation*}
 \mathbf{c}_u^{(t)}\leftarrow
 \left(\sum_{v}\Bb^{(t)}_{v,u}\mathbf{r}_v^{(t)}\right)^{-1},
 \mathbf{r}_v^{(t)}\leftarrow
 \left(\sum_{u}\Bb^{(t)}_{v,u}\mathbf{c}_u^{(t)}\right)^{-1},
\end{equation*}
The second update uses the newly computed column values, and the final vectors become $\mathbf{r}^{(t+1)}$ and $\mathbf{c}^{(t+1)}$. 

We consider three SK scheduling policies: \textsc{Fixed}, \textsc{Adaptive}, and \textsc{Local}. For each of the policies, the outer scaling alone supplies the first proposal. After each step, the \textsc{Fixed} policy performs exactly $K$ full iterations on the residual. After selecting $(v,u)$ at depth $t$, \textsc{Adaptive} performs $\lceil(1-\bar z^{(t)})K\rceil$ iterations, where $z^{(t)}=\mathbf r_v^{(t)}\mathbf c_u^{(t)}$ and $\bar z^{(t)}=\min\{1,\max\{0,z^{(t)}\}\}$. This keeps the budget between zero and $K$. For exactly scaled weighted inputs, the balance loss is $1-\Bb^{(t)}_{v,u}z^{(t)}$. We retain raw $z^{(t)}$ in the tested controllers; it depends on the weight scale, so the weighted results apply to the reported range.

The \textsc{Local} policy replaces the global child iterations with a symmetric
local repair around both removed endpoints.  Let $(v,u)$ be the selected nonzero,
let $\mathcal{N}_t(v)\setminus\{u\}$ and $\mathcal{N}_t(u)\setminus\{v\}$ be the still-active
neighbors of the removed row and column, and define
\begin{small}
\begin{align*}
 \mathcal{C}_{\rm loc}&=\bigl(\mathcal{N}_t(v)\setminus\{u\}\bigr)
   \cup \mathcal{N}_{t+1}\!\bigl(\mathcal{N}_t(u)\setminus\{v\}\bigr),\\
 \mathcal{R}_{\rm loc}&=\bigl(\mathcal{N}_t(u)\setminus\{v\}\bigr)
   \cup \mathcal{N}_{t+1}\!\bigl(\mathcal{N}_t(v)\setminus\{u\}\bigr).
\end{align*}
\end{small}
\noindent Here $\mathcal{N}_{t+1}(U)$ denotes the union of active child neighbors of the vertices in $U$. Using the current, pre-repair scaling vectors, the policy computes
\begin{equation}
 \begin{aligned}
 \mathbf{r}_i^{\rm new}
 &=\left(\sum_{j\in \mathcal{N}_{t+1}(i)}
 \Bb_{i,j}^{(t+1)}\mathbf{c}_j^{\rm old}\right)^{-1},
 &&i\in \mathcal{R}_{\rm loc},\\
 \mathbf{c}_j^{\rm new}
 &=\left(\sum_{i\in \mathcal{N}_{t+1}(j)}
 \Bb^{(t+1)}_{i,j}\mathbf{r}_i^{\rm old}\right)^{-1},
 &&j\in \mathcal{C}_{\rm loc},
 \end{aligned}
 \label{eq:two-sided-light-sk}
\end{equation}
and then replaces both scaling vectors simultaneously. This Jacobi-style update is symmetric in the
removed row and column, and touches only their two-hop child neighborhood, unlike an SK iteration, which touches the entire residual.

Let \(m_{t+1}=\operatorname{nnz}(\Bb^{(t+1)})\), and let
\(m_{\mathrm{loc}}^{(t+1)}\) denote the number of active residual
edges incident to \(\mathcal{R}_{\mathrm{loc}}\cup \mathcal{C}_{\mathrm{loc}}\).
Constructing the two-hop neighborhood and performing the local
repair require $\mathcal{O}\!\left(
|\mathcal{R}_{\mathrm{loc}}|+|\mathcal{C}_{\mathrm{loc}}|
+m_{\mathrm{loc}}^{(t+1)}
\right)
$
work. In the worst case,
\(m_{\mathrm{loc}}^{(t+1)}=\Theta(m_{t+1})\), and {\sc Local}
has no asymptotic advantage over a full SK iteration; its advantage
is empirical and arises when the touched neighborhood is substantially
smaller than the residual graph. For matrices with highly skewed or
power-law degree distributions, hub vertices can make the two-hop
neighborhood comparable in size to the full residual. A bounded-work
variant could cap the number of incident edges examined and approximate
the local normalization sums through neighborhood sampling. We leave the design of such a sampled local repair to future work. 

\paragraph{DM scheduling policies ($\mathcal{P}_{DM}$):}
Except for the initial call applied to $\Ab$ and shared by all trajectories, an inner DM call keeps only the supported nonzeros in a child residual. {Like the SK policies, the DM policies use the scale products as heuristic refresh signals, motivated by the binary exact-scaling analysis above.
\begin{description}[leftmargin=1.2em,itemsep=-1ex,partopsep=1ex,parsep=1ex]
\item[{\sc All}:] Run DM on every nonempty child residual. Hence, the next proposal is normalized over exactly the retained neighbors. The initial matrix and every surviving residual have total support, which yields
$p_{\rm match}=1$. This policy, combined with $\mathcal{P}_{SK}$ = {\sc Fixed}, makes Algorithm~\ref{alg:adaptive-sis} equivalent to Dufoss\'e et al.'s estimator~\cite{DufosseKayaPanagiotasUcar2022} in terms of SK and DM scheduling. 

\item[{\sc{Thrs}}($\theta_{\rm th}$):] Run DM on the residual if
$z^{(t)}=\mathbf{r}^{(t)}_v\mathbf{c}^{(t)}_u$ 
is less than a given {\em threshold} $\theta_{\rm th}$. The support test above motivates this heuristic.
Unlike $\mathcal{P}_{DM}$ = {\sc All}, this policy does not guarantee $p_{\rm match}=1$ since unsupported nonzeros may exist in the child residual.  

\item[{\sc{Trck}}($\theta_{\rm tr}$):] Accumulate the scale-product refresh signal during a trajectory by initializing a variable $track_{\rm dm}=1$ and updating it with 
$track_{\rm dm} \leftarrow track_{\rm dm} \times z^{(t)}$. 
Run DM when $track_{\rm dm}<\theta_{\rm tr}$ and reset
$track_{\rm dm}\leftarrow1$ after every triggered call. The idea is that the product accumulates after consecutive choices, so even several individually large values of $z^{(t)}$ can collectively request a support refresh. 
\end{description}

\section{Experimental Results}\label{sec:experiments}

We examine three questions: 
\begin{itemize}
 \item  \emph{First}, how much do scaling,
two-sided MD, and DM filtering contribute relative to their
cost? 
\item \emph{Second}, which combined SK/DM policy gives the best precision per unit
time as order, sparsity, and numerical weights change?
\item  \emph{Third}, how well do our estimators perform against DeepNIS and across independent runs?
\end{itemize}  

All five variants tested use two-sided MD, $K_0=100$, $K=10$, and a shared
initial DM reduction.  The \texttt{fixed\_all} variant combines fixed SK
with DM at every child.  The \texttt{adaptive\_threshold} and
\texttt{local\_threshold} variants trigger DM by the current scale
product and use adaptive or local SK, respectively; the corresponding tracked
variants accumulate the scale products between DM calls. Both DM thresholds, $\theta_{\rm th}$ and $\theta_{\rm tr}$, are set to $0.33$. Although this yields a nice performance-accuracy tradeoff, it is not fine tuned; alternative thresholds or scaling budgets can be better. 

\paragraph{Matrices used.} 
Constructing a variant benchmark that combines exactly known permanents with
nontrivial estimation difficulty is not straightforward. Exact product formulas are known for the number of perfect matchings in
rectangular grid graphs~\cite{Kasteleyn1961,TemperleyFisher1961}, while
matrices of fixed bandwidth admit exact algorithms whose running time is
linear in the matrix order, with a bandwidth-dependent
constant~\cite{LundowMarkstrom2022,Schwartz2009}. These families, however, provide limited discrimination among the policies considered here. In our
setting, the grid instances also considered by Dufoss\'e~et~al.~\cite{DufosseKayaPanagiotasUcar2022} are handled effectively by the
two-sided minimum-degree rule, whereas regular banded matrices tend to induce
nearly symmetric local choices and, consequently, weakly differentiated
proposal probabilities. Both families are therefore easy for
the estimator. The direct DeepNIS comparison below uses small sparse matrices for which exact reference permanents can still be computed, as given in Table~\ref{tab:suitesparse-matrices} and used for the direct comparison in
Sec.~\ref{sec:deepnis-comparison}. At this scale, however, the
minimum-degree rule often yields a high \(p_{\mathrm{match}}\), as
also observed in the preliminary experiments. 
\begin{table}[pos=htbp]
\centering
\caption{\small Nine real-life matrices used in the comparison.}
\label{tab:suitesparse-matrices}
\setlength{\tabcolsep}{3pt}
\renewcommand{\arraystretch}{0.9}
\scalebox{0.88}{
\begin{tabular}{@{}lrr@{\hspace{10pt}}lrr@{\hspace{10pt}}lrr@{}}
Matrix & $n$ & $m$ & Matrix & $n$ & $m$ & Matrix & $n$ & $m$ \\
\midrule
\texttt{GD95\_c}      & 62 & 287 &
\texttt{bcspwr01}     & 39 & 131 &
\texttt{bcspwr02}     & 49 & 167 \\
\texttt{chesapeake}   & 39 & 340 &
\texttt{curtis54}     & 54 & 291 &
\texttt{dwt\_59}      & 59 & 267 \\
\texttt{ibm32}        & 32 & 126 &
\texttt{mycielskian6} & 47 & 472 &
\texttt{will57}       & 57 & 281 \\
\end{tabular}
}
\end{table}

For a more systematic study of estimation accuracy, we therefore generate
\(50\) randomly permuted block-diagonal matrices with
$n\in\{100,250,500,1000,2000\}$.
For every matrix order and every deletion probability $
p\in\{0.1,0.3,0.5,0.7,0.9\}$, we generate one binary and one weighted instance, giving
\(5\times 5\times 2=50\) matrices in total. Before permutation, each instance
consists of full diagonal blocks of order \(20\) or \(25\), chosen so that
their orders sum to \(n\). Within each block, every non-diagonal entry is
deleted independently with probability \(p\), while all main diagonal entries are
retained. Retaining the diagonal guarantees at least one perfect matching in
every block and hence a positive permanent for the complete matrix. 
Each block permanent is computed exactly using Ryser and multiplied to find the permanent of the block-diagonal matrix. Note that simply adding
superfluous nonzeros to the generated matrices does not necessarily produce a more informative benchmark. If interblock entries are added only in one triangular direction, the resulting matrix remains block triangular: its permanent is unchanged,
and the added edges belong to no perfect matching and are therefore removed
by the initial DM reduction. If interblock entries are added in both
directions, they can create cross-block perfect matchings. In that case, the
block-product formula is lost, and computing the exact permanent at the
target matrix orders is no longer practical.

\begin{figure*}[pos=t]
\centering
    \begin{subfigure}[t]{0.49\textwidth}
        \centering
        \includegraphics[width=\linewidth]{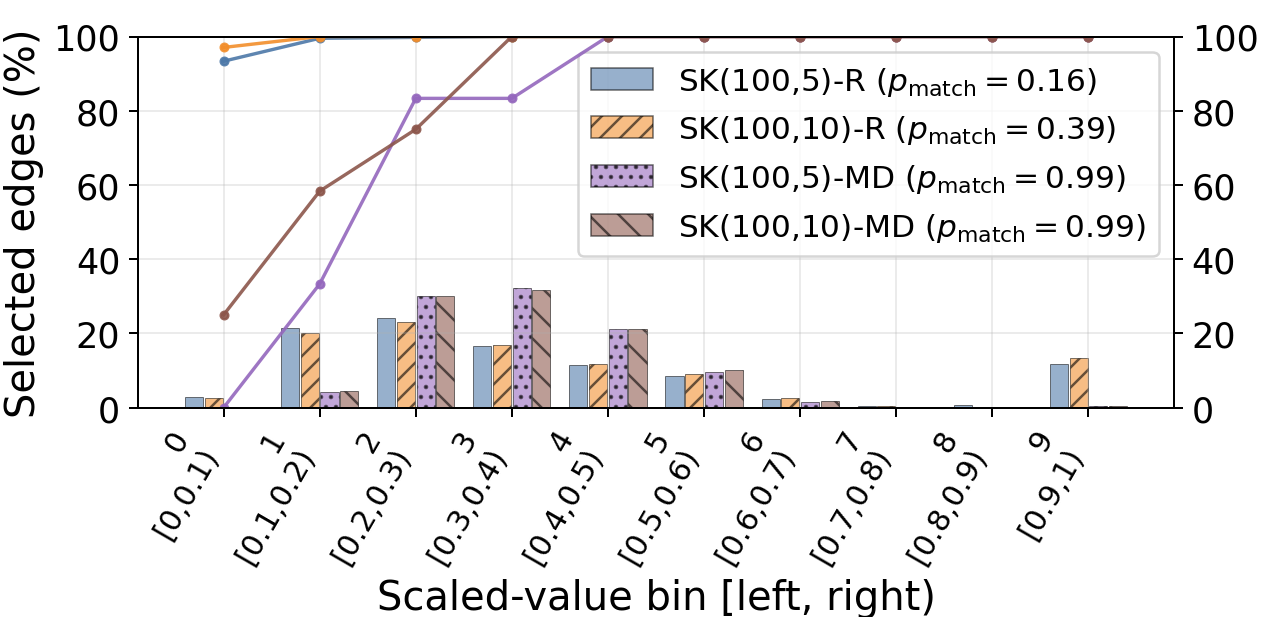}
        \caption{\texttt{sphere3}: $n=258$ and $m=1{,}026$.}
        \label{fig:scaling-fault-sphere3}
    \end{subfigure}
    \begin{subfigure}[t]{0.49\textwidth}
        \centering
        \includegraphics[width=1.0\linewidth]{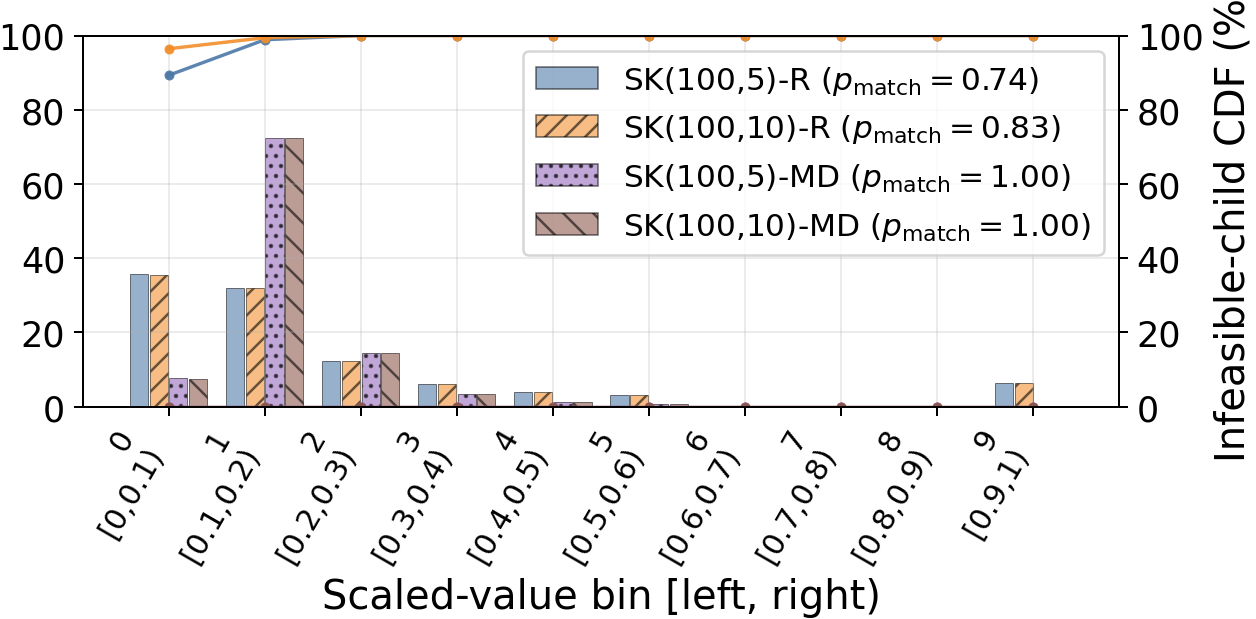}
        \caption{\texttt{Trefethen\_500}: $n=500$ and
        $m=8,478$.}
        \label{fig:scaling-fault-trefethen}
    \end{subfigure}
    \caption{\small{Relation between the scaled value of a selected nonzero and loss of support in its child residual, measured over 1,000}
    trajectories per variant on binary support.  The interval $[0,1]$ is
    divided into ten equal-width buckets. Bars show the
    fraction of selected edges in each bucket, and the right-axis line is the cumulative distribution function (CDF) of selections producing child residuals with no support.
    The legend value $p_{\rm match}$ is the empirical trajectory-completion probability. 
    Failures concentrate in the lower-value buckets, so the scaled value is informative as a DM scheduling signal.}
    \label{fig:scaling-fault-bins}
\end{figure*}

Although connected-component detection readily exposes the blocks and their permanent factorization, the
importance weight of a trajectory is accumulated over the entire matrix.
Consequently, proposal imbalances arising in different blocks can compound
multiplicatively in a single trajectory. As the matrix order and the number
of blocks increase, this construction can therefore amplify
importance-weight dispersion and expose meaningful differences among the
policies, even though the individual block permanents remain exactly
computable. We have used this as a controlled stress test of whole-matrix trajectories; it does not compare against estimating components separately or establish the same behavior on large connected instances. The weighted instances use support realizations generated
independently of the corresponding binary instances, and every retained
entry is assigned a positive value uniformly chosen from \([1,1.5]\). 
We also conduct an unknown-permanent study using the four matrices in Table~\ref{tab:unknown-consensus}.  

\paragraph{Metrics.}
When a reference value \(P\) is known, the relative error of an estimate
\(\widehat P\) is \( |\widehat P-P|/|P| \). For a target tolerance
\(\epsilon\), we define the empirical time to target as the first logged
checkpoint following the last checkpoint at which the relative error exceeds
\(\epsilon\); equivalently, every subsequent logged checkpoint in that run
remains within \(\epsilon\). If no violation occurs, the first logged
checkpoint is used. This retrospective metric uses the exact permanent and later checkpoints; it is neither an online stopping rule nor an accuracy certificate.
We also report \(p_{\rm match}\), the {\em relative spread}
\(s_X/|\overline X|\), and the {\em estimated relative standard error} {(RSE)}
$\frac{s_X}{\sqrt{M}\,|\overline X|}$,
where
\begin{equation*}
s_X=\operatorname{sd}(X_1,\ldots,X_M)
:=\sqrt{\frac{1}{M-1}\sum_{k=1}^{M}(X_k-\overline X)^2}.
\end{equation*}
Spread and RSE are undefined when $M<2$ or $\overline X=0$; an all-zero run is reported as producing no completed trajectory.

We additionally report the sample count and the numbers of DM, full-SK,
and local-SK operations. All statistics include zero-valued trajectory
contributions.

\paragraph{Implementation and platform.}
The estimators are single-threaded C++17 programs compiled with {\tt gcc 11.4} using {\tt -O3} as the optimization level. The server has four 2.30~GHz Intel Xeon E7-4870~v2 sockets, 60 cores, and 503 {GB} of memory. Each estimator run is limited to 3{,}600 secs., while DeepNIS is capped at 10{,}800 secs. per error target.
The runtime profiles exclude the shared initial DM and outer SK phase.
Matching, SCC, edge-marking, and bucket workspaces are allocated once. Scaling
vectors use double precision, while trajectory products and accumulated
statistics use extended precision.

\subsection{SK, MD, and DM}

The ablation studies presented as preliminary experiments (Figs.~\ref{fig:scaling-minor-proxy} and~\ref{fig:md-toolstudy}) identify two complementary effects. 
\begin{enumerate}
\item SK improves the proposal: after ten iterations, the scaled values have correlations close to $0.99$ with the exact child-permanent ratios. 
\item MD
improves the trajectory itself:  With the same outer scaling, fixed row order
produces no completed trajectory on either matrix, whereas
switching only the vertex rule to MD raises $p_{\rm match}$ to
$0.937$ on \texttt{sphere3} and $0.998$ on \texttt{Trefethen\_500}. With 10
inner SK iterations, MD also reduces the relative spread from $4.14$ to $1.68$ and
from $1.84$ to $0.67$.
\end{enumerate}

The important point is that this structural gain is cheap.  MD bucket
maintenance costs only $0.151$ and $0.471$ ms per trajectory on the two
matrices.  A single full inner SK sweep is about $17$--$39\times$ more
expensive, and all-step DM is about $27$--$57\times$ more expensive.  The
incremental implementation therefore makes two-sided MD a practical default:
It removes a major source of dead trajectories without consuming the sampling
budget needed for independent repetitions.

Figure~\ref{fig:scaling-fault-bins} explains why DM should instead be
scheduled.  Choices leading to infeasible children are concentrated at small
scale products, so the scale product is useful for directing structural work.
We do not claim that it is a (total) support certificate; however, this supports our adaptive estimator design: keep the inexpensive MD rule active, use scaling to maintain proposal quality, and reserve DM for risky steps.

\begin{table}[pos=htbp]
\centering
\caption{\small{Results over known permanents of 50 block-diagonal matrices.
``Error wins'' and ``RSE wins'' count matrices on which a variant has the
smallest final absolute relative error or RSE. $M$ is the number of
trajectories. An error-stabilization time~(EST) entry reports ``matrices stabilized/50; median
seconds''; computed with stabilization.}}
\label{tab:block-policy-summary}
\scriptsize
\renewcommand{\arraystretch}{0.9}
\setlength{\tabcolsep}{1.2pt}
\scalebox{1.1}{
\begin{tabular}{@{}lrr@{\hspace{5pt}}rrr@{\hspace{5pt}}r@{}}
Variant & \shortstack{Err.\\wins} & \shortstack{RSE\\wins} &
\shortstack{Median\\RSE} & \shortstack{Min.\\$p_{\rm match}$} &
\shortstack{Median\\$M$} &
\shortstack{EST $10^{-3}$\\(reached; s)} \\
\midrule
\texttt{fixed\_all}          & 3  & 0  & $3.81\times10^{-3}$ & 1.0000 & 40.3K   & 14; 1173.8 \\
\texttt{adaptive\_thr.} & 8  & 1  & $2.27\times10^{-3}$ & 0.9858 & 136.6K   & 25; 1641.6 \\
\texttt{local\_thr.}    & 20 & 42 & $8.09\times10^{-4}$ & 0.9549 & 1.455M & 31; 507.5 \\
\texttt{adaptive\_trc.}   & 5  & 0  & $2.58\times10^{-3}$ & 0.9997 & 105.4K   & 19; 414.6 \\
\texttt{local\_trc.}      & 14 & 7  & $1.40\times10^{-3}$ & 0.9997 & 519.4K   & 27; 767.2 \\
\end{tabular}
}
\end{table}

\subsection{Variant Selection}
\label{sec:block-policy-scaling}

\begin{figure*}[pos=t]
  \centering
  \includegraphics[width=0.93\textwidth]{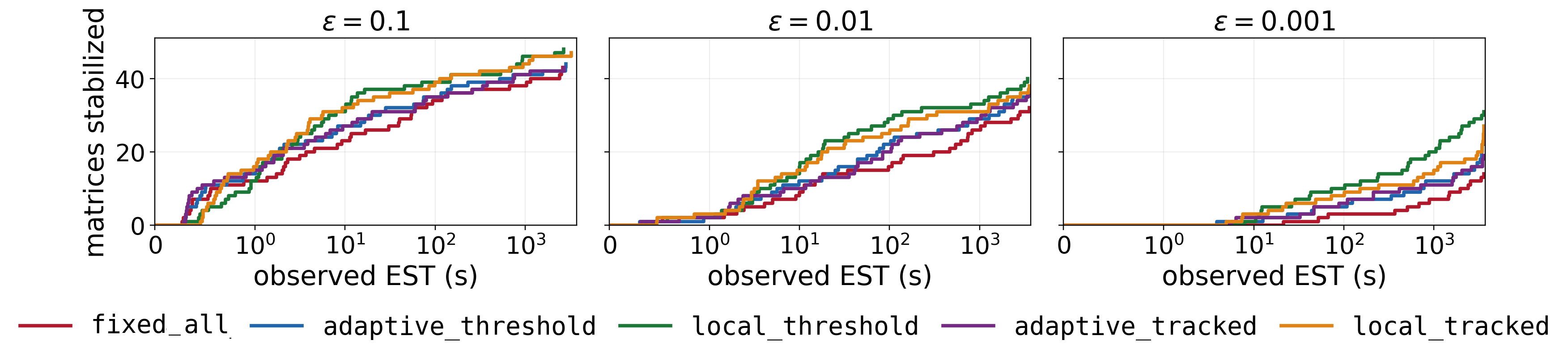}
  \caption{\small{Error-stabilization profiles on the 50
  block-diagonal matrices. A matrix enters a curve only after its final violation of the displayed target. Plateaus show coverage as well as speed.}}
  \label{fig:block-policy-stabilization}
\end{figure*}
The results of the 50-matrix study in Table~\ref{tab:block-policy-summary} show the advantage of using
\texttt{local\_threshold}.  It has the smallest reported RSE on 42 of the 50
matrices and the smallest realized final error on 20. It reaches $10^{-2}$ on 40 matrices
with a conditional median of 17.2 seconds, and reaches $10^{-3}$ on 31 with a
median of 507.5 seconds. A more detailed stabilization profile comparison is given in Figure~\ref{fig:block-policy-stabilization} for $\epsilon \in \{0.1, 0.01, 0.001\}$. Furthermore, in pairwise RSE
comparisons presented in Table~\ref{tab:block-policy-pairwise}, it beats every alternative on at least 42 matrices and beats
\texttt{fixed\_all} on all 50. The contrast with \texttt{fixed\_all} is especially informative. $\mathcal{P}_{DM} = $ {\sc All} guarantees $p_{\rm match}=1$, but it wins no RSE comparison and
produces a median of only $4.03\times10^4$ trajectories. The selective {\sc Local}
policy produces $1.45\times10^6$, about $36\times$ as many, while retaining a
minimum $p_{\rm match}$ of $0.95$.  Hence, certifying every child for total support is
more expensive than losing a small fraction of trajectories. $\mathcal{P}_{SK} = $ {\sc Local}
preserves enough proposal quality, and threshold DM spends the expensive
structural work only where the scale product signals risk. Although $\mathcal{P}_{DM} = $ {\sc Trck} is promising, it is more conservative and invokes DM too often to
match that precision-throughput balance. We leave the sensitivity of {\sc Trck} to $\theta_{\rm{tr}}$ as a future study. 

\begin{figure*}[pos=t]
  \centering
  \includegraphics[width=0.93\textwidth]{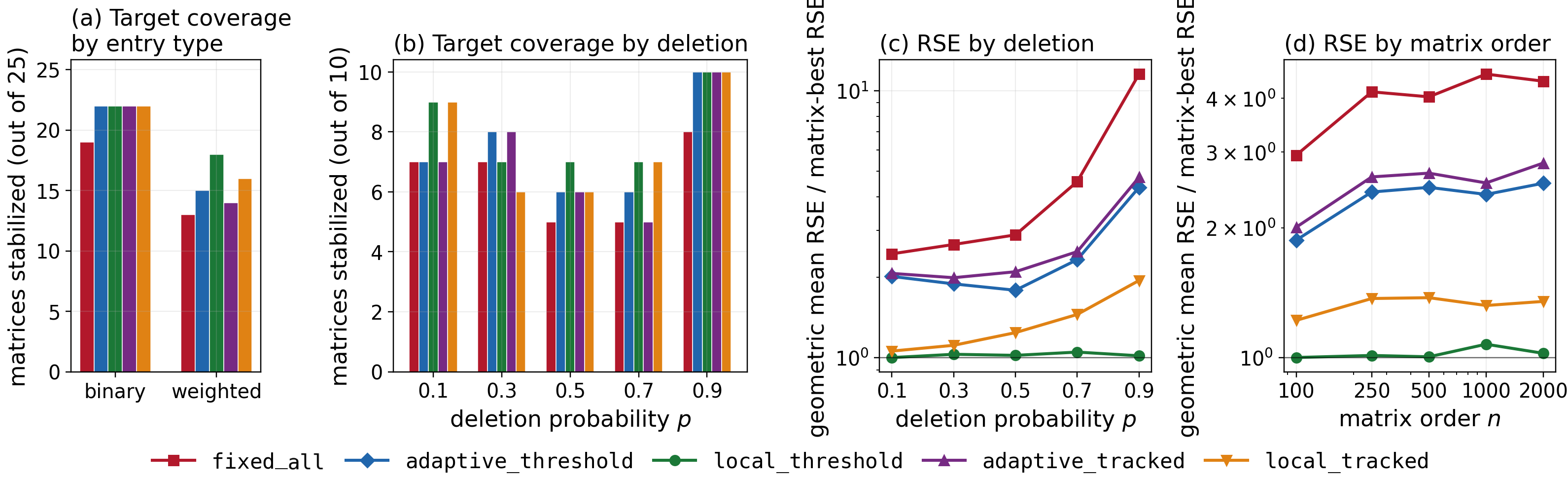}
  \caption{\small{(a) Comparison of binary and weighted coverage at relative error $10^{-2}$: a matrix is counted only if its estimate remains within the target after its final violation. (b) The same coverage by deletion
  probability; a larger $p$ implies more sparsity. (c) and (d): The geometric mean RSE relative to the best variant on each matrix.}}
  \label{fig:block-policy-summary}
\end{figure*}

\begin{table}[pos=htbp]
\caption{\small{
Pairwise variant wins over the 50 block-diagonal matrices.
Each entry \(a/b\) gives the number of matrices on which row variant \(A\)
has a strictly smaller final absolute relative error (\(a\)) and RSE
(\(b\)) than column variant \(B\). For either metric, opposite
off-diagonal counts sum to 50.}}
\label{tab:block-policy-pairwise}
\centering
\footnotesize
\setlength{\tabcolsep}{2.0pt}
\renewcommand{\arraystretch}{0.92}
\scalebox{0.98}{
\begin{tabular}{@{}lccccc@{}}
\toprule
                    & {\texttt{fixed}} & {\texttt{adapt.}} & {\texttt{local}}  & {\texttt{adapt.}} & {\texttt{local}}  \\
\(A\backslash B\)   & {\texttt{all}} & {\texttt{thresh.}} & {\texttt{thresh.}} &  {\texttt{tracked}} & {\texttt{tracked}} \\
\midrule
\texttt{fixed\_all}  & --      & \(11/0\)  & \(6/0\)   & \(14/1\)  & \(7/0\)  \\
\texttt{adaptive\_threshold}  & \(39/50\) & --      & \(13/2\)  & \(23/45\) & \(21/1\) \\
\texttt{local\_threshold}   & \(44/50\) & \(37/48\) & --      & \(41/49\) & \(29/42\) \\
\texttt{adaptive\_tracked}  & \(36/49\) & \(27/5\)  & \(9/1\)   & --       & \(13/0\) \\
\texttt{local\_tracked}   & \(43/50\) & \(29/49\) & \(21/8\)  & \(37/50\) & --       \\
\bottomrule
\end{tabular}}
\end{table}

\begin{table}[pos=htbp]
\caption{\small Numbers of binary/weighted matrices stabilized at relative
error $10^{-2}$, out of 5 matrices at each order.}
\label{tab:stabilization-by-order}
\centering
\footnotesize
\setlength{\tabcolsep}{4pt}
\renewcommand{\arraystretch}{0.87}

\begin{tabular}{@{}rccccc@{}}
\toprule
                    & {\texttt{fixed}} & {\texttt{adapt.}} & {\texttt{local}}  & {\texttt{adapt.}} & {\texttt{local}}  \\
$n$  & {\texttt{all}} & {\texttt{thresh.}} & {\texttt{thresh.}} &  {\texttt{tracked}} & {\texttt{tracked}} \\
\midrule
100  & $5/5$ & $5/5$ & $5/5$ & $5/5$ & $5/5$ \\
250  & $5/5$ & $5/5$ & $5/5$ & $5/5$ & $5/5$ \\
500  & $5/1$ & $5/2$ & $5/4$ & $5/2$ & $5/3$ \\
1000 & $2/1$ & $4/2$ & $5/2$ & $4/1$ & $5/2$ \\
2000 & $2/1$ & $3/1$ & $2/2$ & $3/1$ & $2/1$ \\
\bottomrule
\end{tabular}
\end{table}

Figure~\ref{fig:block-policy-summary}(a) shows that the estimators stabilize on more binary instances than weighted instances, while Figure~\ref{fig:block-policy-summary}(b) shows lower coverage at the intermediate deletion probabilities. The weighted instances exhibit greater observed dispersion among positive trajectory products; for \texttt{local\_threshold}, their geometric mean RSE is $4.35\times$ the binary value. Still, as Table~\ref{tab:stabilization-by-order} shows, the variants with ${\mathcal P}_{SK} =$ {\sc Local} stabilize for $n \in \{100, 250\}$ on all five weighted instances, whereas {\tt local\_threshold} also stabilizes on four of five for $n = 500$. For the latter, we hypothesize that dense blocks offer balanced alternatives, while at $p = 0.9$, most entries are removed by DM calls.

Figure~\ref{fig:block-policy-summary}(c) and (d) show that compared to {\tt fixed\_all}, the benefits of work filtering become more pronounced as the deletion probability $p$ and matrix order $n$ increase. The marginal value of unconditional DM decreases sharply as the matrices become sparser. The median threshold-DM rate falls by almost three orders of
magnitude from $p=0.1$ to $p=0.9$, while the trajectory-count advantage of
\texttt{local\_threshold} over \texttt{fixed\_all} grows from about $6\times$
to $180\times$. Although increasing $n$ reduces absolute precision (because each trajectory is longer and fewer trajectories fit in the time budget), it does not change the ranking: \texttt{local\_threshold} has the smallest RSE on seven to nine of the ten matrices at each tested order.

\subsection{Comparison with DeepNIS}\label{sec:deepnis-comparison}

DeepNIS~\cite{HarviainenKoivisto2024} is run separately at the requested $\epsilon\in\{0.1,0.01,0.001\}$ with failure-probability parameter $\delta=0.05$ and initial DM enabled. DeepNIS supplies an $(\epsilon,\delta)$ accuracy statement, whereas our timed samplers are evaluated by realized error and retrospective stabilization under different time caps; the comparison does not equate these guarantees. 

A DeepNIS run is counted as {\em valid} only if it returns an estimate in $10{,}800$ seconds with an error meeting the target. For this experiment, 9 matrices from Table~\ref{tab:suitesparse-matrices} and 10 $(100 \times 100)$ block-diagonal matrices from the previous experiment are used. Under the reported time caps, DeepNIS covers fewer of the larger test instances than the proposed variants.  
Figure~\ref{fig:deepnis-profiles} shows that the coverage gap widens as the requested accuracy tightens. At $\epsilon=0.01$, all five variants stabilize on all 19 matrices, while DeepNIS satisfies the target on 10. At $\epsilon=0.001$,
\texttt{local\_threshold} still covers all 19; the other variants
cover 17 or 18, and DeepNIS has four valid outputs. On the block matrices, DeepNIS outputs at $\epsilon=0.001$ only for the sparse and dense extremes; for the weighted instances, it does not meet the target even at those extremes.

Because a target-driven method can exceed its requested accuracy, the preceding comparison may disadvantage DeepNIS. We also
match outcomes: for each valid DeepNIS output with a realized error $\widehat{\epsilon}$, we measure when a sampler first stabilizes at $\widehat{\epsilon}$.
Table~\ref{tab:deepnis-outcome} reports this comparison for the two tightest
requested targets. The reached counts in Figure~\ref{fig:deepnis-profiles}
remain essential because each conditional median uses only outputs for which
the corresponding sampler also reaches $\widehat{\epsilon}$.

\begin{figure}[pos=htbp]
  \centering
  \includegraphics[width=0.98\linewidth]{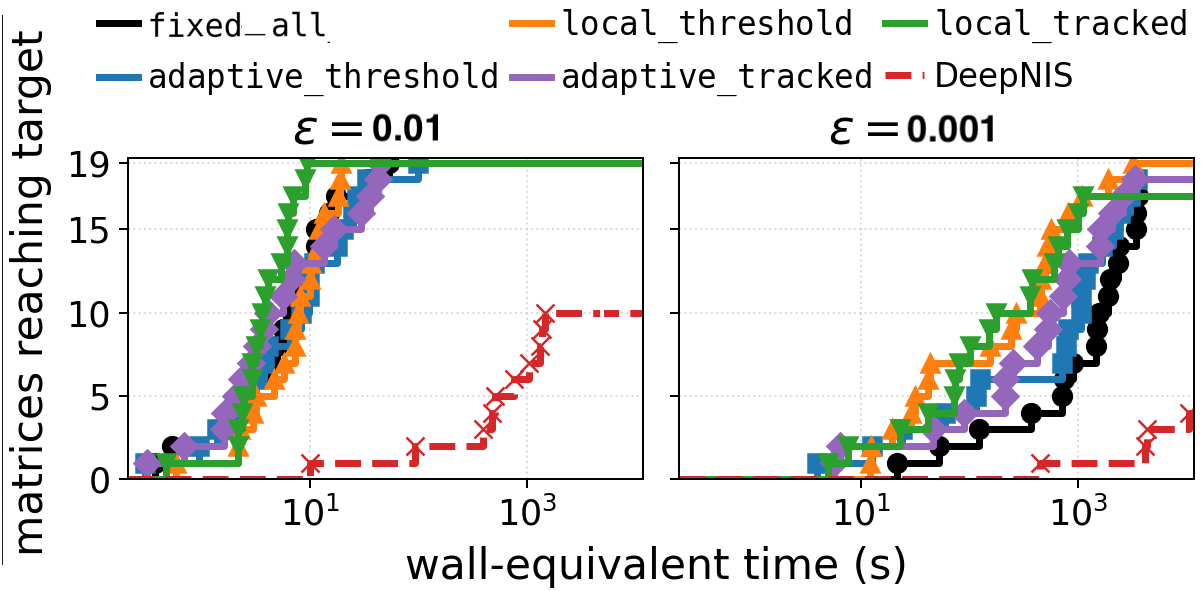}
  \caption{\small{Accuracy-target profiles for 9 real-life and
ten order-100 block matrices. Estimator curves use stabilization
times from 3{,}600-second traces.  Each DeepNIS point uses the run requested at
the same target with a 10{,}800 second timeout.}}
  \label{fig:deepnis-profiles}
\end{figure}

\begin{table}[pos=htbp]
\centering
\caption{\small{
Outcome-matched times for DeepNIS runs with
\(\epsilon=10^{-2}\) and \(10^{-3}\).
The estimator and DeepNIS columns report conditional median times in
seconds. The final column is the median paired estimator-to-DeepNIS
time ratio; a value below one favors the proposed estimator variant.
For \(\epsilon=10^{-2}\), DeepNIS meets the target for
10 instances; for \(\epsilon=10^{-3}\), DeepNIS meets the target for
4.}}
\label{tab:deepnis-outcome}

\footnotesize
\setlength{\tabcolsep}{2.5pt}
\renewcommand{\arraystretch}{0.90}

\begin{tabular}{@{}clrrr@{}}
\toprule
&
Variant &
\shortstack{Estimator\\time (s)} &
\shortstack{DeepNIS\\time (s)} &
\shortstack{Time\\ratio} \\
\midrule

\multirow[c]{5}{*}{
  \rotatebox[origin=c]{90}{\(\epsilon=0.01\)}
}
& \texttt{fixed\_all}          & 409.3 & 376.6 & 1.711 \\
& \texttt{adaptive\_threshold} & 64.4  & 386.2 & 0.464 \\
& \texttt{local\_threshold}    & 41.5  & 427.8 & 0.470 \\
& \texttt{adaptive\_tracked}   & 211.9 & 386.2 & 0.549 \\
& \texttt{local\_tracked}      & 30.7  & 376.6 & 0.576 \\

\cmidrule(lr){2-5}

\multirow[c]{5}{*}{
  \rotatebox[origin=c]{90}{\(\epsilon=0.001\)}
}
& \texttt{fixed\_all}          & 2125.0 & 4126.3 & 0.340 \\
& \texttt{adaptive\_threshold} & 225.6  & 4126.3 & 0.506 \\
& \texttt{local\_threshold}    & 63.5   & 4126.3 & 0.142 \\
& \texttt{adaptive\_tracked}   & 95.5   & 2286.2 & 0.174 \\
& \texttt{local\_tracked}      & 31.8   & 2286.2 & 0.073 \\
\bottomrule
\end{tabular}
\end{table}

At $\epsilon=0.01$, each variant except \texttt{fixed\_all} has a median time ratio
below one. At
$\epsilon=0.001$, all ratios are below one, although the
comparison is based on only four DeepNIS runs. The experiment shows that \texttt{local\_threshold} continues to return an empirically accurate estimate across the full test set, while the other proposed configurations retain substantially broader coverage than DeepNIS.

\subsection{Larger Real-World Matrices}\label{sec:unknown-permanents}

The final study uses the four matrices in
Table~\ref{tab:unknown-consensus}. 
Each variant is run for 3{,}600 seconds under
five independent random seeds, resulting in 100 runs in total.
Because the true permanents are unavailable, the analysis
focuses on replication, cross-variant agreement, and observed uncertainty.

Let $Y_{{\mathcal P}, r}$ be the final estimate from the variant with policy-set $\mathcal P$ and seed $r$, with trajectory
standard deviation $s_{{\mathcal P}, r}$ and sample count $M_{{\mathcal P}, r}$.  The within-run standard
error is $e_{{\mathcal P}, r}=s_{{\mathcal P}, r}/\sqrt{M_{{\mathcal P}, r}}$.  For the five seeds of policy set ${\mathcal P}$, define $ \overline Y_{\mathcal P}=\frac{1}{5}\sum_{r=1}^{5}Y_{{\mathcal P},r}$, and
\begin{equation*}
 \begin{aligned}
 e^{\rm in}_{\mathcal P}&=\frac{\sqrt{\sum_r e_{{\mathcal P},r}^2}}{5},&
 e^{\rm seed}_{\mathcal P}&=\frac{ {\operatorname{sd}(Y_{\mathcal{P}, 1}, \ldots, Y_{\mathcal{P}, 5})}}{\sqrt5}.
 \end{aligned}
 \label{eq:unknown-variant-se}
\end{equation*}
We set $e_{\mathcal P}=\max(e^{\rm in}_{\mathcal P},e^{\rm seed}_{\mathcal P})$ and compute the equal-variant
consensus
$\widehat P_{\rm con}=\frac{1}{5}\sum_{\mathcal P}\overline Y_{\mathcal P}$ and $e_{\rm con}={\sqrt{\sum_{\mathcal P} e_{\mathcal P}^2}}/{5}$.
The equal mean is unbiased when its constituent estimates are unbiased, but the displayed interval remains an empirical diagnostic: the stated propagation of errors assumes independent runs across variants, and rare large-weight trajectories may be absent from every run.

\begin{table}[pos=htbp]
\centering
\caption{\small{Five-seed consensus for four unknown 
permanents. The relative half-width, ``Rel. HW'', is $1.96e_{\rm con}/|\widehat P_{\rm con}|$;
``variant ratio'' is the largest variant mean divided by the smallest; and
``min ESS'' is the smallest run-level approximate effective sample size.}}
\label{tab:unknown-consensus}
\footnotesize
\setlength{\tabcolsep}{2pt}
\renewcommand{\arraystretch}{0.9}
\scalebox{0.93}{
\begin{tabular}{lrrrrrrl}
\toprule
 & &  &  &  Variant  &  & \\
Matrix & $n$ & $m$ & Rel. HW & ratio & Min ESS & Status\\
\midrule
\texttt{Trefethen\_500} & 500 & $8,478$ & $1.83\mathrm{E}{-3}$ & 1.001 & 10,849 & provisional\\
\texttt{dwt\_503}       & 503 & 6,027  & $5.16\mathrm{E}{-3}$ & 1.013 & 3,819  & provisional\\
\texttt{olm500}         & 500 & 1,996  & $9.01\mathrm{E}{-4}$ & 1.001 & 33,032 & provisional\\
\texttt{tomography}     & 500 & 28,726  & $9.10\mathrm{E}{-2}$ & 1.242 & 5      & unresolved\\
\bottomrule
\end{tabular}}
\end{table}

For the cross-variant consistency diagnostic, define
\begin{equation*}
D_{\max}
=
\max_{\mathcal{P} < \mathcal{P'}}\left(
{\left|\overline{Y}_\mathcal{P}-\overline{Y}_\mathcal{P'}\right|}
     /{\sqrt{e_\mathcal{P}^2+e_{\mathcal{P}'}^2}}\right).
\end{equation*}
The five variants induce \(\binom{5}{2}=10\) pairwise
comparisons. We use a two-sided Bonferroni correction at family-wise
level \(\alpha=0.05\), yielding the critical value
\begin{equation*}
c_{\mathrm{Bonf}}
=
\Phi^{-1}\!\left(
1-\frac{\alpha}{2\binom{5}{2}}
\right)
=
\Phi^{-1}(0.9975)
\approx 2.81,
\end{equation*}
where \(\Phi\) is the standard-normal distribution function. A matrix
passes this test when \(D_{\max}\leq c_{\mathrm{Bonf}}\), i.e.,
that none of the pairwise differences is detected as significant under the Bonferroni-adjusted normal approximation.

We label a matrix \emph{provisional consensus} only when all 25 runs
are present, the five variant-wise 95\% \(t\)-intervals have a
common intersection, \(D_{\max}\leq c_{\mathrm{Bonf}}\), every run has at least 1,000 positive trajectories, an approximate effective sample size of at least 1,000, and the approximate consensus 95\% half-width
\(1.96e_{\mathrm{con}}\) is at most 10\% of
\(\lvert\widehat P_{\mathrm{con}}\rvert\). Three of the four matrices satisfy every diagnostic. 
Their variant means differ by at most $1.29\%$, and their relative consensus half-widths are below $0.52\%$.
The \texttt{tomography} instance is qualitatively different. Every variant completes every
trajectory, yet the mean RSEs range from $0.125$ to $0.261$, and the
minimum ESS is only five. This case makes our central distinction explicit:
Eliminating dead ends does not control heavy dispersion among positive contributions. In practice, more work is necessary; independent repetitions should be checked together with ESS, RSE, and the largest contribution to the sample sum to detect unresolved states. 

The variant comparison favors selective local work; across the 20 matrix--seed pairs, \texttt{local\_threshold} has the smallest RSE in 13,
compared with six for \texttt{local\_tracked} and one for
\texttt{adaptive\_tracked}.  It also generates $62.851$ million trajectories,
more than $18\times$ the next-highest variant total, and has the lowest median
run-level RSE.  \texttt{local\_tracked} is marginally better only on
\texttt{Trefethen\_500}. Thus \texttt{local\_threshold} remains the strongest
empirical precision-per-hour variant.

\section{Related Work}\label{sec:related-work}

{\em{Exact and certified approximation.}}
Ryser--Glynn inclusion--exclusion and modern parallel implementations remain
the standard route to exact permanents, but their exponential dependence on
$n$ limits the attainable order~\cite{ElbekTasyaranUcarKaya2026,Glynn2010,LundowMarkstrom2022,Ryser1963}.  Randomized approximation can
provide stronger guarantees: Karmarkar~et~al. give a mildly exponential Monte
Carlo method for binary permanents~\cite{doi:10.1137/0222021},
Jerrum and Sinclair reduce approximate counting to sampling matchings
~\cite{doi:10.1137/0218077}, and Jerrum, Sinclair, and Vigoda give a {\em fully polynomial randomized approximation scheme} (FPRAS) for
all nonnegative matrices~\cite{JerrumSinclairVigoda2004}.  Later annealing
improvements reduce the polynomial cost~\cite{Bezakova}, although available
implementations remain expensive in practice~\cite{Newman}.  These methods
supply worst-case probabilistic guarantees that our estimator does
not.

\paragraph{Scaling and sequential estimators.}
Matrix scaling supports both deterministic bounds and proposal construction.
Linial~et~al. use strongly polynomial scaling for
permanental bounds~\cite{LinialSamorodnitskyWigderson2000}; related bounds for
doubly stochastic matrices and the Bethe permanent provide efficiently
computable approximations with exponential-factor guarantees
~\cite{AnariRezaei2019,Vontobel2013}. Sullivan and Beichl sample
from SK-balanced, dense matrices~\cite{SullivanBeichl2014}, whereas
Dufoss\'e et al. combine scaling with repeated removal of unsupported nonzeros for sparse matrices~\cite{DufosseKayaPanagiotasUcar2022}.

\paragraph{Rejection and nesting.}
Self-reducible rejection methods use permanent bounds to control acceptance~\cite{HuberLaw2008}; AdaPart adapts the partition
~\cite{KuckDaoRezatofighiSabharwalErmon2019}, {and deep acceptance-rejection evaluates bounds deeper in the recursion tree~\cite{HarviainenRoyskoKoivisto2021}.  Subsequent nesting methods strengthen the bounds or sample in proportion to them~\cite{HarviainenKoivisto2024}, and their ability to attach an $(\epsilon,\delta)$ statement to an estimate is important. 

The present work combines these ideas under a different cost objective. It
maintains exact residual degrees on both bipartite sides with
$\mathcal{O}(n + m)$ total bucket work per trajectory, rather than recomputing it, as in the prior work discussed above. It decouples vertex selection, proposal repair, and structural filtering: cheap two-sided MD keeps trajectories active, local or adaptive SK controls proposal quality, and {DM is invoked when the scale-product trigger signals a risky step. This separation
preserves conditional unbiasedness for weighted nonnegative matrices while
allowing a small loss in $p_{\rm match}$ when the saved work yields many more samples per hour.

\section{Conclusion and Future Work}\label{sec:conclusion}

We design an unbiased sequential estimator for sparse nonnegative
matrix permanents by separating three decisions that are often coupled:
which residual row or column to expand, how much scaling work to perform, and
when to pay for exact structural filtering. 
The experiments show that selective work can be more valuable than the
certification of support.  Although all-step DM guarantees $p_{\rm match}=1$, it wins no
RSE comparison on the 50-matrix benchmark because its per-trajectory cost
severely limits sample count.  By contrast, \texttt{local\_threshold} combines local scaling with scale-triggered DM and obtains the smallest RSE on 42 of 50 matrices. The advantage does not come from maximizing completion; it comes from allocating the time budget more effectively among proposal repair, structural filtering, and independent repetitions.

Against DeepNIS, the proposed variants have broader empirical coverage as the accuracy target tightens.  At $10^{-3}$, \texttt{local\_threshold} stabilizes
on all 19 known permanents, while DeepNIS meets the target on only four under the
reported caps. This empirical comparison obviously does not replace DeepNIS's $(\epsilon,\delta)$ guarantee. However, as a future work, we are planning to apply the bounds in the literature to our estimator.

\FloatBarrier
\appendix
\section{{Bounds supporting the schedules}}
\label{app:scheduling-bounds}

\paragraph{A sharper scaling bound.}
Let $\Bb$ be a binary matrix of order $k\geq2$, with exact doubly stochastic scaling $\Sb=\diag(\mathbf r)\Bb\diag(\mathbf c)$. For a selected nonzero, put $z=\mathbf r_v\mathbf c_u$ and $U(\Bb)=(\prod_i\mathbf r_i\prod_j\mathbf c_j)^{-1}$. Let $\widetilde\Bb$ be the DM-filtered child, and let $\rho_i,\kappa_j$ be its row and column sums under the inherited factors. If $\delta$ is the scaled mass removed by DM within the child, then
\begin{equation}
 \begin{aligned}
 \eta&=\min\left\{\prod_i\rho_i,\prod_j\kappa_j\right\},\\
 \perm(\Bb\setminus\{v,u\})&\leq U(\widetilde\Bb)
                 \leq z\eta U(\Bb),
 \end{aligned}
 \label{eq:scaling-bound-reduction}
\end{equation}
\begin{equation}
 \eta\leq\left(1-\frac{1-z+\delta}{k-1}\right)^{k-1}
       \leq\exp(z-1-\delta).
 \label{eq:scaling-factor-refinement}
\end{equation}
To prove~\eqref{eq:scaling-bound-reduction}, row-normalize the inherited child factors in the scaling objective of~\cite{DufosseKayaPanagiotasUcar2022}. This gives $U(\widetilde\Bb)\leq zU(\Bb)\prod_i\rho_i$. Column normalization gives the other product. Their sums equal $k-2+z-\delta$, so the inequality of arithmetic and geometric means yields~\eqref{eq:scaling-factor-refinement}. Thus $\eta$ sharpens the factor $\exp(z-1)$ in Lemma~2~\cite{DufosseKayaPanagiotasUcar2022}. Note that in practice, a larger $\delta$ records more unsupported mass removed from the child and tightens the bound. This observation can also be used to implement a better SK schedule which we plan to pursue in the future. 

For the binary sampler with exact scaling and DM at every step, the bounds telescope:
\begin{equation*}
 \begin{aligned}
 X(\sigma)&\leq U(\Ab)\prod_t\eta_t(\sigma),\\
 \E[X^2]&\leq\perm(\Ab)U(\Ab)\,
             \E_\sigma\!\left[\prod_t\eta_t(\sigma)\right].
 \end{aligned}
\end{equation*}
Here vertex selection is deterministic, $\sigma$ is uniform over perfect matchings, and $\eta_t=1$ for $k = 1$. Each matching has sampling probability $1/X(\sigma)$. This replaces $\exp(z_t-1)$ by $\eta_t$ in Theorem~3 of~\cite{DufosseKayaPanagiotasUcar2022}.

For instance, for $\Ab=\mathbf I+\mathbf P_n$, where $\mathbf P_n$ is the
permutation matrix of a single $n$-cycle, we have
$\perm(\Ab)=2$, $U(\Ab)=2^n$, and
$\prod_t\eta_t=2^{-(n-1)}$, since DM leaves a unique matching
after the first selection, i.e., every possible residual matrix has a single perfect matching. 
The new second-moment bound is exact, $\E[X^2]=4$, whereas
Theorem~3 of~\cite{DufosseKayaPanagiotasUcar2022} gives
$2^{n+1}\exp(-1/2)$.

\FloatBarrier
\balance

\section*{Funding}
This research did not receive any specific grant from
funding agencies in the public, commercial, or not-for-profit
sectors.

\section*{Data availability}
The code supporting this study is available
at \url{https://github.com/kamerkaya/the-quick-and-the-dead}.

\section*{{Declaration of generative AI and AI-assisted
technologies in the manuscript preparation process}}

During the preparation of this work, the authors used
ChatGPT (OpenAI) to assist with language editing and literature searches. 
After using this tool, the authors reviewed and edited the
content as needed and take full responsibility for the
content of the published article.

\bibliographystyle{elsarticle-num}
\bibliography{references}

\end{document}